\documentclass{article}
\usepackage{times}
\usepackage{graphics} 
\def\refn{\noindent\hangafter=1\hangindent=1cm}
\input epsf
\begin{document}

\vspace{-1cm}
\centerline{\large\bf A catalog of variable stars in the Kepler field of view}
\centerline{\large\bf found in the ASAS3-North data}
%\centerline{\large\bf The All Sky Automated Survey.}
%\centerline{\large\bf The catalog of variable stars in the Kepler field of view}

\vspace{0.3cm}
\centerline{by}

\vspace{0.3cm}
\centerline{A.\,Pigulski$^1$, G.\,Pojma\'nski$^2$, B.\,Pilecki$^2$, D.\,Szczygie{\l}$^2$}

\vspace{0.3cm}
\noindent{\small $^1$Instytut Astronomiczny Uniwersytetu Wroc{\l }awskiego, Kopernika 11, 51-622 Wroc{\l }aw, Poland}\\
{\small $^2$Warsaw University Astronomical Observatory, Al. Ujazdowskie 4, 00-478 Warszawa, Poland}

\vspace{0.5cm}
\noindent{\bf Introduction}

\vspace{0.5cm}
\noindent Although the launch of Kepler mission (Borucki et al.~1997) is scheduled for February 2009, the asteroseismic targets 
(Christensen-Dalsgaard et al.~1997) for the first year should be proposed as soon as possible. 
In order to facilitate the selection of targets, we have observed the Kepler field of view using the All Sky Automated Survey - North (ASAS3-North) 
instruments for over a year. Then we have performed a detailed search for variability which resulted in detection of almost 1000 variable 
stars. We announce the release of the catalog containing the list of these stars, their properties and finding charts. 

\vspace{5mm}
\noindent{\bf Observations and analysis}

\vspace{3mm}
\noindent The data analyzed here were obtained in the ASAS3-North station located at Haleakala (Maui, Hawaii Islands) using two wide-field instruments, 
equipped with Nikkor 200-mm f/2.0 lenses and Apogee AP-10, 2048$\times$2048 CCD cameras, collecting data in two filters (V and I). The data cover 
roughly 500 days between July 2006 and December 2007. The resolution of ASAS images amounts to almost 15 arcsec/pixel. The details of image 
processing and reductions can be found in the previous ASAS papers (Pojma\'nski 1997, 2002).

The variability search was carried out using the data in the intermediate aperture (dia\-me\-ter of 4 pixels) by means of Fourier amplitude 
periodogram calculated in the range between 0 and 30 d$^{-1}$. All light curves and periodograms were inspected visually. The variability 
type was assigned taking into account the period, amplitude and/or the shape of the light curve. In total, 947 stars were selected as 
variable among about 250,000 searched for variability. Only stars that will be covered by the CCD chips of Kepler satellite were considered. 
The catalog does not contain variable stars that are located in the gaps between chips.

\vspace{5mm}
\noindent{\bf The availaibility and contents of the catalog}

\vspace{3mm}
\noindent The catalog is available at two mirror sites: 
\begin{quote}
{\it http://www.astro.uni.wroc.pl/ldb/asas/kepler.html}
\end{quote}
or
\begin{quote}
{\it http://www.astrouw.edu.pl/asas/?page=kepler}.
\end{quote}

The catalog is presented in a form of a table that includes the most important information on a given star, namely: the identification, the
equatorial coordinates, VI photometry from ASAS, JHK photometry from 2MASS (Cutri et al.~2003), type of variability, 
period (if applicable), ranges of variability in V and I and cross-identifications.  The latter were taken from the other sources of information on
variability, i.e., Combined General Catalog of Variable Stars (Samus et al.~2004), ROTSE1 catalog (Akerlof et al.~2000),
Northern Sky Variability Survey (NSVS, Wo\'zniak et al.~2004a) catalogs of red variables (Wozniak et al.~2004b) and RR Lyrae stars 
(Wils et al.~2006) and Hungarian-made Automated Telescope Network (HATnet) catalog (Hartman et al.~2004) in the field \#199.

The data, finding charts and additional information can be retrieved following a link in the second column of the table, i.e., the 
master ASAS identification of a star.  The link opens a new page. It contains a header, links to the original VI data
and panels showing the VI light curves. The light curves are also shown in phase diagrams provided that the star is a periodic or quasiperiodic 
variable with period shorter than 150 days.  Next, the finding charts from ASAS and Digital Sky Survey are shown. Finally, there is a table
with information on a star and its position in the Kepler field of view.

\vspace{5mm}
\noindent{\bf Notes on variability type and follow-up observations}

\vspace{3mm}
\noindent Eleven variability classes have been assigned to the variable stars presented in the catalog. Three (EA, EW, EB) are used
for eclipsing binaries with light curves typical for Algol, $\beta$~Lyrae and W~UMa-type, respectively, as defined in the General Catalog
of Variable Stars (GCVS). Five other are used for five classes of pulsating variables that can be easily classified using the period(s),
amplitude(s) and the shape of the light curve. These are: Miras (MIRA), Cepheids (CEP), RR Lyrae stars of Bailey type ab (RRAB) and c (RRC)
and high-amplitude $\delta$~Scuti stars (HADS). The remaining three classes (PER, QPER, APER) are used for stars for which the type
of variability cannot be presently assigned. It would be possible if additional information (e.g., spectral type) will become available. 
Stars classified as PER are strictly
periodic with a light curve that could not be distinguished from a simple sinusoid. Stars classified as QPER show dominating periodicity, 
but also amplitude and/or phase changes or a superimposed variability on a longer time scale. Stars of this type are commonly designated 
as semi-regular or long-period variables. Finally, stars designates as APER show no well-defined periodicity in light variation.

The class defined as PER contains potentially many variables which are pulsating stars, but their type cannot be unambiguously given
from the knowledge of only a period as two or more types of variability overlap in period. This is the case of $\delta$~Scuti, $\beta$~Cephei and RRc
stars on one hand, and slowly-pulsating B (SPB) stars and $\gamma$~Doradus stars, on the other.  Next, some non-pulsating stars
like low-amplitude W UMa binaries, ellipsoidal or $\alpha^2$~CVn variables may also be classified as PER, especially because their
amplitude of variability can be significantly reduced due to the contamination by nearby bright stars.  In consequence, the variability
type of some stars classified as PER needs to be verified by the follow-up observations.  We have already started follow-up programs in the 
Bia{\l}k\'ow observatory of Wroc{\l}aw University (Poland) and Konkoly Observatory (Hungary).  Observers interested in verifying variability of
the short-period variable stars from our catalog are asked to write to A.\,Pigulski (pigulski@astro.uni.wroc.pl).

\vspace{0.5cm}
{\bf Acknowledgements.} We are indebet to Mr.\,Wayne Rosing who has kindly allocated space and provided technical support of the LCOGT 
staff for ASAS-North instruments inside The Faulkes Telescope North on Haleakala, Maui, Hawaii.
This project was supported by the N20300731/1328 grant from Polish Ministry of Science.

\vspace{0.5cm}
\centerline{REFERENCES}
\vspace{0.3cm}
{\small
\refn Akerlof C., Amrose S., Balsano R., et al., 2000, AJ 119, 1901.\par
\refn Borucki W.J., Koch D.G., Dunham E.W., Jenkins J.M., 1997, ASPC 119, 153.\par
\refn Christensen-Dalsgaard J., Arentoft T., Brown T.M., et al., 2007, Comm.~in Asteros.~150, 350.\par
\refn Cutri R.M., Skrutskie M.F., Van Dyk S., et al., 2003, {\it 2MASS All-Sky Catalog of Point Sources}.\par
\refn Hartman J.D., Bakos G., Stanek K.Z., Noyes R.W., 2004, AJ 128, 1761.\par
\refn Pojma\'nski G., 1997, Acta Astron.~47, 467.\par
\refn Pojma\'nski G., 2002, Acta Astron.~52, 397.\par
\refn Samus N.N., Durlevich O.V., et al., 2004, {\it Combined General Catalog of Variable Stars} (GCVS 4.2, 2004 Ed.).\par
\refn Wils P., Lloyd C., Bernhard K., 2006, MNRAS 368, 1757.\par
\refn Wo\'zniak P.R., Vestrand W.T., Akerlof C.W., et al., 2004a, AJ 127, 2436.\par
\refn Wo\'zniak P.R., Williams S.J., Vestrand W.T., Gupta V., 2004b, AJ 128, 2965.\par

}

\end{document}